# Reference Model for Performance Management in Service-Oriented Virtual Organization Breeding Environments

Zbigniew Paszkiewicz, Willy Picard

Department of Infromation Technology, Poznan University of Economics,
Mansfelda 4, 60-854 Poznań, Poland
{zpasz, picard}@kti.ue.poznan.pl

**Abstract.** Performance management (PM) is a key function of virtual organization (VO) management. A large set of PM indicators has been proposed and evaluated within the context of virtual breeding environments (VBEs). However, it is currently difficult to describe and select suitable PM indicators because of the lack of a common vocabulary and taxonomies of PM indicators. Therefore, there is a need for a framework unifying concepts in the domain of VO PM. In this paper, a reference model for VO PM is presented in the context of service-oriented VBEs. In the proposed reference model, both a set of terms that could be used to describe key performance indicators, and a set of taxonomies reflecting various aspects of PM are proposed. The proposed reference model is a first attempt and a work in progress that should not be supposed exhaustive.

**Keywords:** virtual organization, performance measurement, key performance indicators

## 1 Introduction

The concept of *Virtual Breeding Environment* (VBE) has been proposed by the ECOLEAD project as "an association of organizations and their related supporting institutions, adhering to a base long term cooperation agreement, and adoption of common operating principles and infrastructures, with the main goal of increasing their preparedness towards collaboration in potential *Virtual Organizations (VO)*" [1]. The main aims of VBE are the establishment of trust for organization to collaborate in VOs, the reduction of cost and time to search and find suitable partners of VOs, the assistance in VO creation, re-configuration, reaching agreement [1]. In ECOLEAD *Virtual organization* is defined in the following manner: "an *operational structure* consisting of *different organizational entities* and created for a *specific business purpose*, to address a *specific business opportunity*" [2].

Among various important aspects of VBE management and VO Management (VOM) identified by the ECOLEAD project, the concept of *VO Performance Measurement (PM)* has be defined as a "systematic approach to plan and conduct the collection and monitoring of data for performance indicators. The Performance

Measurement is focused on the process of data collection. The input are defined Performance Indicators (PI) including targets. For these PIs appropriate methods and tools of measuring have to be developed and planed. The collected data is then prepared into indicators. As a result performance measurement produces reports that could be used for further analyses and interpretation to assess the performance and to derive measures for improvement" [3].

Measuring the performance of distributed organization is a difficult task. Many works have justified the need for PM and have analyzed existing approaches in the management science as regards potential adaptation to VO [3–5]. However the studied approaches usually concentrate on specific aspects of organization operation, e.g. logistics, accomplishment of strategic goals, financial aspects. As a consequence, measuring the performance of a whole VO, and not only some aspects of it, still requires a more global approach. Additionally, to our best knowledge, no systematic classification of existing indicators exists, and therefore, it is difficult to describe and select suitable indicators and sources of data for their calculation.

Therefore, there is a need for a VO-specific *Reference Model (RM)* for VO PM. The goal of such a reference model should be two-fold: on the one hand, the reference model should define a set of common terms that could be used to describe key performance indicators (KPIs). On the second hand, the reference model should define various taxonomies of KPIs, each taxonomy focusing on various aspects of PM.

In this paper, a first attempt of a Reference Model for VO PM is presented. This paper is organized as follows. In Section 2, the concept of Service-Oriented VBEs (SOVOBE) is introduced, as our reference model focuses on such a type of VBE. Then, the importance of the concept of KPIs in context of SOVOBE is explained. In Section 4, a Reference Model for VO PM is proposed. Finally, Section 5 concludes the paper.

## 2  Service-Oriented VBE

### 2.1  SOA and CNOs

While the concept of VBE is currently rather accepted in the Collaborative Network Organization (CNO) research community, there is currently no consensus about the architecture and implementation of VBEs.

However, the Service-Oriented Architecture (SOA) has been suggested as a valuable approach for the architecture and implementations of VBEs in [6]. The definition of SOA by the OASIS group [7] is the following one: "Service Oriented Architecture (SOA) is a paradigm for organizing and utilizing distributed capabilities that may be under the control of different ownership domains. […] in SOA, services are the mechanism by which needs and capabilities are brought together." This definition emphasizes some characteristics of SOA common to CNOs: CNOs may be seen as structures aiming at "organizing and utilizing distributed capabilities under the control of different ownership domains".

### 2.2 Main Elements of SOVOBEs

In this paper, only service-oriented virtual organization breeding environments (SOVOBE) are taken into account. In a SOVOBE, VBE and VO operations are based on services performed by people, organizations and information systems, composed in complex business processes.

Additionally, in this paper, it is assumed that a *social network* is supporting the SOVOBE by providing information about relations among people, organizations, information systems, and business processes.

A social network is a graph of *nodes* (sometimes referred as *actors*), which may be connected by *relations* (sometimes referred as *ties*, *links*, or *edges*). Social Network Analysis (SNA) is the study of these relations [8] and may be used to analyze the structure of SOVOBEs.

An important aspect of SNA is the fact that it focused on the how the structure of relationships affects actors, instead of treating actors as the discrete units of analysis. SNA is backed by social sciences and mathematical theories like graph theory and matrix algebra [9], which makes it applicable to analytical approaches and empirical methods. SNA uses various concepts to evaluate different network properties.

## 3 KPIs in SOVOBE

### 3.1 KPIs and SLAs in SOVOBE

*Key Performance Indicator (KPI)* has been defined in the ECOLEAD project as "a performance indicator which represents essential or critical components of the overall performance. Not every performance indicator is vital for the overall success of a company. To focus attention and resources the indicators with the highest impact on performance are chosen as key performance indicators. […] An *indicator* is a variable which is feasible to assess the state of an object in scope. Indicators could be as well quantitative as qualitative measures. They can not only consist of a single measure but also be aggregated or calculated out of several measures" [3].

A similar concept has been proposed in SOA: *Service Level Agreement (SLA)*. A SLA is a part of a service contract where the level of service is formally defined. A SLA usually refers to a single service and a single organizational unit that is responsible for provision of this service.

The difference between KPIs and SLAs in SOVOBE is related with the scope of these concepts.

**Table 1.** Scope of SLAs and KPIs in SOVOBE

| SLA | KPI |
|---|---|
| Service | - |
| Process | Composition of services within a process |
| Partner | Composition of services within a partner |
|  | Composition of processes within a partner |
| VO | Composition of processes within VO |
| | Composition of partners within VO |
| VBE | Composition of VO within VBE |

The scope of interest of SLAs and KPIs in SOVOBE is presented in Table 1. KPIs concentrate on compositions of elements of SOVOBE. For instance, the subject of a KPI is not a performance of a single service but a composition of services. Similarly, a KPI does not measure the performance of a single process, but of a composition of processes. On the other hand, composition of services may be perceived from process point of view – process as a composition of services – or partner point of view – partner as a composition of services that are provided by him. In turn, process may be analyzed in context of one partner – process a partner takes part in – or VO – processes that VO operation is based on. Finally VO is a composition of partners and VBE is a composition of VO. Again, it is important to stress, that KPIs defined on a level of single SOVOBE element (i.e. partner) do not refer to the element itself but to collection of its components.

### 3.2 Anticipation and Monitoring in CNOs

KPIs may be used through almost entire CNO life cycle. Especially, they are useful in CNO creation and operational phase. The possible set of KPIs that might be used in mentioned phases is the same, but the approach to the measurement changes.

**Anticipation.** During CNO creation, the aim of PM is to *anticipate* the CNO overall performance. A set of KPIs, together with expected values, are defining performance requirements for the CNO to be created. Then by comparing expected values of KPIs to calculated values for a chosen potential CNO realization, a CNO planner may identify strengths and weaknesses of the chosen potential CNO realization.

**Monitoring** During CNO operation, constant monitoring of KPIs may take place. While anticipation is active, monitoring is passive. The performance of existing composition of artifacts is evaluated using KPIs and the result of the evaluation is compared with performance requirements. If there are any deviations from the accepted level of performance of a partner, process or a whole network, CNO can take actions to replace service, partner, process, information system or change business goal.

## 4  A Reference Model for KPIs in SOVOBE

The OASIS group have defined the concept of reference model as follows: "A reference model is an abstract *framework* for understanding significant relationships among the entities of some environment. It enables the development of specific reference or concrete architectures using consistent standards or specifications supporting that environment. A reference model consists of a minimal set of *unifying concepts, axioms* and relationships within a particular problem domain, and is *independent* of specific standards, technologies, implementations, or other concrete details" [7].

The main contribution of this paper is a reference model for KPIs in SOVOBE. In this section, both a set of terms that could be used to describe KPIs, and a set of taxonomies reflecting various aspects of PM are proposed. The proposed reference model is a first attempt and a work in progress that should not be supposed exhaustive.

### 4.1 Data Source

A significant issue related with KPIs is the accessibility of data. Some KPIs can be computed using publicly available data stored in SOVOBE, e.g. history of collaboration, description of services, opinion of services. However, the calculation of other KPIs requires access to data stored in partners' internal databases, usually not accessible for technical or organizational reasons (organization may not allow public access to certain piece of data, nonetheless this access may be granted as a result of negotiations). A potential solution to this problem consists in accessing the data via services. Such a solution implies an agreement among partners on conditions of providing additional services. This agreement might be reached during the process of partner selection and negotiation. These additional services may be composed in a *control process* that will be probably synchronized with the main *operational process* ruling partners' interaction.

Therefore, the following KPI subcategories presented in Table 2 may be distinguished:

**Table 2.** KPI Data source

|       | **Category name** | **Description** | **Example** |
|-------|-------------------|-----------------|-------------|
| 1.    | Collaboration-based | Based on a data strictly connected with provision of services needed for operational process | |
| 1.1   | Subjective | Coming form a subjective opinion of one of parties involved in collaboration | |
| 1.1.1 | Service consumer opinion | Data provided by a service consumer, based on his perception of reality | Perceived time of partner's response |
| 1.1.2 | Service provider opinion | Data provided by service provider | Partner guarantees referring failure rate of services |
| 1.2   | Objective | Not dependent on opinion of parties involved in collaboration | |
| 1.2.1 | Continuous monitoring of collaboration | Data provided by monitoring of service use and process progress | Current time of partner response, current partner's reliability |

|  | Category name | Description | Example |
|---|---|---|---|
| 1.2.2 | Bag of assets | Data stored in VBE or VO | |
| 1.2.2.1 | History of collaboration | Data restored form the history of partners' performance and collaboration within VBE | Number of VOs a partner participated in, partner's average failure rate |
| 1.2.2.2 | Description of services | Quantitative values hold in service description | Formally declared time of response |
| 1.2.2.3 | Description of competences | Quantitative values hold in description of partner and competences | Number of services offered by partner |
| 1.2.2.4 | Contracts and SLA | Agreed conditions of cooperation | Real cost of the service in a particular process |
| 1.2.3 | Social network | Data modeled in SN, data coming from third parties that are not directly involved in evaluated collaboration process | Experience of the partner, acknowledgement of the partner |
| 2. | Non-collaboration-based | Based on additionally negotiated data not required in a operational process | |
| 2.1 | Control process | Data accessible within control process | Personal data of partner's subcontractors |

The proposed subcategories are not exclusive, as the calculation of a KPI may require various sources.

### 4.2 Subject of Measurement

As mentioned in Section 2, it is possible to distinguish following elements of a SOVOBE: service, process, partner, VO, and VBE. As described in section 3.1 single services are not a subject of measurement by KPIs. Table 3 presents subcategories referring to consecutive elements.

**Table 3.** KPI subject of measurement

|  | Category name | Description | Example |
|---|---|---|---|
| 1. | Process | Composition of services | Total cost of the process calculated on a basis of service costs |
| 2. | Partner | Composition of services and processes | Partner's reliability calculated on a basis of average failure rate of services provided by partner |
| 3. | Virtual Organization | Composition of processes and partners | Number of partners involved in more than one VO |
| 4. | Virtual Breeding Environment | Composition of VOs | Average number of partners in VO |

The proposed subcategories are not exclusive, as the calculation of a KPI may concerns various subjects. A KPI could determine the importance of a particular partner the ratio of a number of services provided by a partner to a total number of services used in a process.

### 4.3 Scope

KPIs may be defined at various level of granularity. The most typical case is to define a KPI for a given VO, with the KPI measuring performance aspects related with the specific characteristics of particular VO. However, some KPIs may be defined at the VBE level and be shared by all VOs, with KPI measuring performance aspects related with the specific characteristics of particular VBE [10].

KPIs defined at the VBE level would allow e.g. VO comparison, uniform validation of quality of created networks of cooperation, imposition of best practices and validation of conformation to these practices.

**Table 4.** KPI scope

|    | Category name | Description |
| --- | --- | --- |
| 1. | Global | KPIs obligatory for all VOs despite the VBE |
| 2. | Standard | KPIs obligatory for all VO within a given VBE |
| 3. | Custom | KPIs defined for a particular VO within particular VBE |

Identified KPI scopes are presented in Table 4.

### 4.4 Performance of Collaboration

In SOVOBE, performance of collaboration is conditioned by *structure of the network* and by quality of SOVOBE elements. KPIs referring to the structure of collaboration are called *structural indicators* and their definition and analysis is based on SNA. Structure of the network represents relations among artifacts. Structure is modeled in a social network. Among all multiple views and aspects of the network structure for performance analysis, it is useful to distinguish the structure of services. S*tructure of services* is a structure of a graph representing composition, dependencies, and complexity of services. Relations among services directly influence fulfillment of an operational process and influence the operation of VO and VBE.

A second important aspect in performance measurement is a quality of SOVOBE elements and non-functional requirements that can be defined for them. KPIs measuring the use and operations are referred as *operational indicators*. Operational indicators are presented e.g. in the ISO 9126 standard [11] and in the ECOLEAD project [3]. A taxonomy of KPIs related with the performance of collaboration is presented in Table 5.

**Table 5.** KPI for performance of collaboration

|     | Category name | Description | Example |
| --- | --- | --- | --- |
| 1.  | Structural | Referring to a structure of collaboration network | |
| 1.1 | Service structure | Referring to a structure of collaboration (service composition) directly influencing operational process | Number of service provided by a partner for the network, degree of VO overlapping |
| 1.2 | General structure | Referring to critical for VO aspects different to a structure of collaboration i.e. partners' experience, competences, acknowledgment etc. | Level of trust, number of VO an organization is involved, level of experience |
| 2.  | Operational | Referring to the quality and non-functional requirements of partners', processes, and offered services | |
| 2.1 | Effectiveness | Reliability of the service and ability to meet expectations | Failure rate |
| 2.2 | Flexibility | Maximal additional capacity that could be provided to the VO | Possible additional production volume in comparison with contracted (%) |
| 2.3 | Substitutability | Ease of replacement of an SOVOBE element | Number of partners with the same competences |
| 2.4 | Efficiency | Usage of resources | Number of involved partners in a process |
| 2.5 | Responsiveness | Time of response | Time to fulfill the request |
| 2.6 | Cost | Cost of operation | Cost of VO process, cost of VO service |
| 2.7 | Productivity | Volume of requests that could be fulfilled | Number of offered services, number of products offered by VO |

## 5   Conclusions

The main contribution presented in this paper is a definition of a Reference Model for KPIs in SOVOBEs. The proposed reference model is based on results of the ECOLEAD project [3–5] concerning performance management. The proposed Reference model aims at providing the CNO community with a vocabulary and a set of taxonomies useful to describe KPIs. However, the Reference Model presented in this paper is a work in progress and it may not be assumed that it is exhaustive.

Among future works, the reference model should be extended by practitioners that may confront the proposed reference model with the needs of CNO members in terms of performance management. It would also be interesting to confront the proposed reference model with the reference model for Collaborative Networks ARCON proposed by Camarinha-Matos et al. [2].

Finally, within the context of the IT-SOA project [12], a service-oriented VBE is currently under development. This VBE will gather companies from the construction sector in the Great Poland region. During the IT-SOA project, the Key Performance Indicators will be defined in a real business environment, based on an extended version of the presented Reference Model.

**Acknowledgments.** This work has been partially supported by the Polish Ministry of Science and Higher Education within the European Regional Development Fund, Grant No. POIG.01.03.01-00-008/08.